\documentclass[ALICE,manyauthors]{cernphprep}
\usepackage{hyperref}
\usepackage{lineno}
\usepackage{xcolor}
\usepackage{euscript}
\usepackage{cite}
\usepackage[scientific-notation=true]{siunitx}
\usepackage[greek,english]{babel}
\usepackage{multirow}

\usepackage[Symbolsmallscale]{upgreek}
\usepackage[T1]{fontenc}
\usepackage{orcidlink}

\begin{document}
%

\newcommand{\pp}           {pp\xspace}
\newcommand{\ppbar}        {\mbox{$\mathrm {p--\overline{p}}$}\xspace}
\newcommand{\XeXe}         {\mbox{Xe--Xe}\xspace}
\newcommand{\PbPb}         {\mbox{Pb--Pb}\xspace}
\newcommand{\pAa}           {\mbox{pA}\xspace}
\newcommand{\pPb}          {\mbox{p--Pb}\xspace}
\newcommand{\AuAu}         {\mbox{Au--Au}\xspace}
\newcommand{\dAu}          {\mbox{d--Au}\xspace}
\newcommand{\pP}{\ensuremath{\mbox{p--p}}\,}


\newcommand{\kstar}        {\ensuremath{k^*}\xspace}
\newcommand{\rstar}     {\ensuremath{r^{*}}\xspace}

\newcommand{\snn}          {\ensuremath{\sqrt{s_{\mathrm{NN}}}}\xspace}
\newcommand{\pt}           {\ensuremath{p_{\rm T}}\xspace}
\newcommand{\meanpt}       {$\langle p_{\mathrm{T}}\rangle$\xspace}
\newcommand{\ycms}         {\ensuremath{y_{\rm CMS}}\xspace}
\newcommand{\ylab}         {\ensuremath{y_{\rm lab}}\xspace}
\newcommand{\etarange}[1]  {\mbox{$\left | \eta \right |~<~#1$}}
\newcommand{\yrange}[1]    {\mbox{$\left | y \right |~<~#1$}}
\newcommand{\dndy}         {\ensuremath{\mathrm{d}N_\mathrm{ch}/\mathrm{d}y}\xspace}
\newcommand{\dndeta}       {\ensuremath{\mathrm{d}N_\mathrm{ch}/\mathrm{d}\eta}\xspace}
\newcommand{\avdndeta}     {\ensuremath{\langle\dndeta\rangle}\xspace}
\newcommand{\dNdy}         {\ensuremath{\mathrm{d}N_\mathrm{ch}/\mathrm{d}y}\xspace}
\newcommand{\Npart}        {\ensuremath{N_\mathrm{part}}\xspace}
\newcommand{\Ncoll}        {\ensuremath{N_\mathrm{coll}}\xspace}
\newcommand{\dEdx}         {\ensuremath{\textrm{d}E/\textrm{d}x}\xspace}
\newcommand{\RpPb}         {\ensuremath{R_{\rm pPb}}\xspace}

\newcommand{\nineH}        {$\sqrt{s}~=~0.9$~Te\kern-.1emV\xspace}
\newcommand{\seven}        {$\sqrt{s}~=~7$~Te\kern-.1emV\xspace}
\newcommand{\onethree}        {$\sqrt{s}~=~13$~Te\kern-.1emV\xspace}
\newcommand{\twoH}         {$\sqrt{s}~=~0.2$~Te\kern-.1emV\xspace}
\newcommand{\twosevensix}  {$\sqrt{s}~=~2.76$~Te\kern-.1emV\xspace}
\newcommand{\five}         {$\sqrt{s}~=~5.02$~Te\kern-.1emV\xspace}
\newcommand{\twosevensixnn}{$\sqrt{s_{\mathrm{NN}}}~=~2.76$~Te\kern-.1emV\xspace}
\newcommand{\fivenn}       {$\sqrt{s_{\mathrm{NN}}}~=~5.02$~Te\kern-.1emV\xspace}
\newcommand{\LT}           {L{\'e}vy-Tsallis\xspace}
\newcommand{\GeVc}         {Ge\kern-.1emV$/c$\xspace}
\newcommand{\MeVc}         {Me\kern-.1emV$/c$\xspace}
\newcommand{\GeVmass}      {Ge\kern-.1emV$/c^2$\xspace}
\newcommand{\MeVmass}      {Me\kern-.1emV$/c^2$\xspace}
\newcommand{\lumi}         {\ensuremath{\mathcal{L}}\xspace}

\newcommand{\ITS}          {\rm{ITS}\xspace}
\newcommand{\TOF}          {\rm{TOF}\xspace}
\newcommand{\ZDC}          {\rm{ZDC}\xspace}
\newcommand{\ZDCs}         {\rm{ZDCs}\xspace}
\newcommand{\ZNA}          {\rm{ZNA}\xspace}
\newcommand{\ZNC}          {\rm{ZNC}\xspace}
\newcommand{\SPD}          {\rm{SPD}\xspace}
\newcommand{\SDD}          {\rm{SDD}\xspace}
\newcommand{\SSD}          {\rm{SSD}\xspace}
\newcommand{\TPC}          {\rm{TPC}\xspace}
\newcommand{\TRD}          {\rm{TRD}\xspace}
\newcommand{\VZERO}        {\rm{V0}\xspace}
\newcommand{\VZEROA}       {\rm{V0A}\xspace}
\newcommand{\VZEROC}       {\rm{V0C}\xspace}
\newcommand{\Vdecay} 	   {\ensuremath{V^{0}}\xspace}

\newcommand{\ee}           {\ensuremath{e^{+}e^{-}}} 
\newcommand{\pip}          {\ensuremath{\pi^{+}}\xspace}
\newcommand{\pim}          {\ensuremath{\pi^{-}}\xspace}
\newcommand{\kap}          {\ensuremath{\rm{K}^{+}}\xspace}
\newcommand{\kam}          {\ensuremath{\rm{K}^{-}}}
\newcommand{\pbar}         {\ensuremath{\rm\overline{p}}\xspace}
\newcommand{\kzeros}        {\ensuremath{{\rm K}^{0}_{\rm{S}}}\xspace}
\newcommand{\kzerobar}     {\ensuremath{\rm \overline{K}^0}}
\newcommand{\lmb}          {\ensuremath{\Lambda}\xspace}
\newcommand{\almb}         {\ensuremath{\overline{\Lambda}}\xspace}
\newcommand{\prot}         {\ensuremath{\rm{p}}\xspace}
\newcommand{\aprot}         {\ensuremath{\rm{\overline{p}}}\xspace}
\newcommand{\n}         {\ensuremath{\rm{n}}\xspace}
\newcommand{\an}         {\ensuremath{\rm{\overline{n}}}\xspace}
\newcommand{\kbar}         {\ensuremath{\rm\overline{K}}\xspace}

\newcommand{\Om}           {\ensuremath{\Omega^-}\xspace}
\newcommand{\Mo}           {\ensuremath{\overline{\Omega}^+}\xspace}
\newcommand{\X}            {\ensuremath{\Xi^-}\xspace}
\newcommand{\Ix}           {\ensuremath{\overline{\Xi}^+}\xspace}
\newcommand{\Xis}          {\ensuremath{\Xi^{\pm}}\xspace}
\newcommand{\Oms}          {\ensuremath{\Omega^{\pm}}\xspace}
\newcommand{\SigZ}            {\ensuremath{\Sigma^0}\xspace}
\newcommand{\aSigZ}            {\ensuremath{\overline{\Sigma^0}}\xspace}
\newcommand{\antik}   {$\mathrm{\overline{K}}\,$}

\newcommand{\Ledn}         {Lednick\'y--Lyuboshits\xspace}
\newcommand{\chiEFT}       {\ensuremath{\chi}\rm{EFT}\xspace}
\newcommand{\ks}     {\ensuremath{k^{*}}\xspace}
\newcommand{\rs}     {\ensuremath{r^{*}}\xspace}
\newcommand{\mt}     {\ensuremath{m_{\mathrm{T}}}\xspace}
\newcommand{\Cth}           {C_\mathrm{th}\xspace}
\newcommand{\Cexp}           {C_\mathrm{exp}\xspace}
\newcommand{\CF}           {\ensuremath{C(\ks)}\xspace}
\newcommand{\Sr}            {\ensuremath{S(\rs)}\xspace}
\newcommand{\BBar}            {\ensuremath{\rm{B}\mbox{--}\rm{\overline{B}}}\xspace}
\newcommand{\SPi}         {\ensuremath{\uppi\Sigma}\xspace}
\newcommand{\kbarN}         {\ensuremath{\rm\overline{K}N}\xspace}
\newcommand{\LL}            {\ensuremath{\lmb\mbox{--}\lmb}\xspace}
\newcommand{\pprot}            {\ensuremath{\prot\mbox{--}\prot}\xspace}

\newcommand{\LK}{$\Lambda \mbox{--}{\mathrm{K}}$\xspace} 
\newcommand{\LAK}{$\Lambda \mbox{--}{\mathrm{\overline{K}}}$\xspace} 
\newcommand{\LKzero}{$\Lambda \mbox{--}{\mathrm{K^0 _S}}$\xspace} 

\newcommand{\LKInt}{$\Lambda{\mathrm{K}}$\xspace} 
\newcommand{\LAKInt}{$\Lambda{\mathrm{\overline{K}}}$\xspace}
\newcommand{\LKMinInt}{$\Lambda$K$^-$\,}
\newcommand{\LKPlusInt}{$\Lambda$K$^+$\,}

\newcommand{\LKPlus}{$\Lambda \mbox{--}$K$^+$\,}
\newcommand{\ALKMin}{$\overline{\Lambda} \mbox{--}$K$^-$\,}
\newcommand{\LKMin}{$\Lambda \mbox{--}$K$^-$\,}
\newcommand{\ALKPlus}{$\overline{\Lambda} \mbox{--}$K$^+$\,}
\newcommand{\LKpair}{$\Lambda \mbox{--} $K$^+\oplus \overline{\Lambda} \mbox{--} $K$^-$\xspace}
\newcommand{\ALKpair}{$\Lambda \mbox{--} $K$^-\oplus \overline{\Lambda} \mbox{--} $K$^+$\xspace}
\newcommand{\KpMin}{K$^-p$\xspace}

\newcommand{\SAK}{$\Sigma\mathrm{\overline{K}}$\xspace} 
\newcommand{\SK}{$\Sigma \mathrm{K}$\xspace}

\newcommand{\Imscatt}            {\ensuremath{\Im f_0}\xspace}
\newcommand{\Rescatt}            {\ensuremath{\Re f_0}\xspace}
\newcommand{\effran}            {\ensuremath{d_0}\xspace}

\newcommand{\ImscattAmpl}            {\ensuremath{\Im f}\xspace}
\newcommand{\RescattAmpl}            {\ensuremath{\Re f}\xspace}

\newcommand{\XRes}          {\ensuremath{\Xi\mathrm{(1620)}}\xspace}
\newcommand{\XResNovanta}          {\ensuremath{\Xi\mathrm{(1690)}}\xspace}
\newcommand{\XResVenti}          {\ensuremath{\Xi\mathrm{(1820)}}\xspace}

\begin{titlepage}
\PHyear{2023}       
\PHnumber{106}      
\PHdate{30 May}     
\title{Accessing the strong interaction between $\Lambda$ baryons and charged kaons with the femtoscopy technique at the LHC}
\ShortTitle{}   

\Collaboration{ALICE Collaboration\thanks{See Appendix~\ref{app:collab} for the list of collaboration members}}
\ShortAuthor{ALICE Collaboration} 

\begin{abstract}
The interaction between \lmb baryons and kaons/antikaons is a crucial ingredient for the strangeness $S=0$ and $S=-2$ sector of the meson--baryon interaction at low energies. In particular, the \LAKInt might help in understanding the origin of states such as the \XRes, whose nature and properties are still under debate. Experimental data on \LK and \LAK systems are scarce, leading to large uncertainties and tension between the available theoretical predictions constrained by such data. In this Letter we present the measurements of \LKpair and \ALKpair correlations obtained in the high-multiplicity triggered data sample in \pp collisions at \onethree recorded by ALICE at the LHC. The correlation function for both pairs is modeled using the \Ledn analytical formula and the corresponding scattering parameters are extracted. The \ALKpair correlations show the presence of several structures at relative momenta \kstar above 200 \MeVc, compatible with the $\Omega$ baryon, the \XResNovanta, and \XResVenti resonances decaying into \LKMin pairs. The low \kstar region in the \ALKpair also exhibits the presence of the \XRes state, expected to strongly couple to the measured pair. 
The presented data allow to access the \LKPlusInt and \LKMinInt strong interaction with an unprecedented precision and deliver the first experimental observation of the \XRes decaying into \LKMinInt. 
\end{abstract}
\end{titlepage}

\setcounter{page}{2} 

\section{Introduction}
Measurements of correlations of particle pairs in the relative momentum space performed in small colliding systems at the Large Hadron Collider (LHC), such as proton--proton (\pp) and \pPb collisions, have proven to be a sensitive experimental tool to investigate hadron--hadron interactions. In recent years, this so-called femtoscopy technique~\cite{Lisa:2005dd} was employed by the ALICE Collaboration to deliver a large amount of high-precision data on interactions involving strange baryons and antibaryons~\cite{ALICE:Run1,ALICE:pXi,ALICE:pSig0,ALICE:pOmega,ALICE:pL,ALICE:LL,ALICE:LXi,Vale_BBar}. This made it possible to validate for the first time state-of-the-art lattice QCD predictions at the physical point and to provide crucial experimental tests for low-energy effective field theories.
Lately, the same technique was applied to meson--baryon pairs giving the possibility to access the interaction of protons with $\phi$, charm D mesons, and kaons~\cite{ALICE:pphi,ALICE:pD,ALICE:pKpp,ALICE:pKCoupled}.
In the strangeness $S=-1$ meson--baryon sector, the measurement of the \kam p correlation function in different colliding systems~\cite{ALICE:pKpp,ALICE:pKPb,ALICE:pKCoupled} provided a detailed picture of the \kam p strong interaction above threshold and novel constraints on the coupling strength to the \kzerobar n and \SPi channels. These femtoscopic correlations delivered the most precise data on the \kam p interaction and a crucial input to pin down the \kbarN--\SPi dynamics, responsible for the formation of the $\Lambda(1405)$ resonance~\cite{Hall:2014uca,Kamiya:2015aea,Kamiya:2016oao}, which currently is the only accepted molecular state in the hadronic spectrum.

States with a similar nature, namely arising dynamically in multi-channel interactions, are predicted to exist also in the $S=-2$ meson--baryon sector in which antikaons (\kbar) interact with the strange $\Lambda$ baryon. Theoretical calculations based on chiral unitary frameworks~\cite{ALK_Ramos,Garcia-Recio:ChiPtXi,Sekihara:LKComp,Miyahara:Xi,ALK_Hyodo,ALK_HyodoPaper}, Bethe-Salpeter approaches~\cite{Wang:BetheSalp}, and meson-exchange models~\cite{Chen:XiOBE} indicate that several $\Xi$ resonances, such as the \XRes and the \XResNovanta, might indeed originate from the coupling between the \LAK system and other $S=-2$ channels, like $\uppi \Xi$ and \SAK. The knowledge on these low-lying $\Xi$ resonances is rather scarce. Several measurements are available~\cite{Ross:Xi1620_1Exp,Briefel:Xi1620_2Exp,BelleXi,BESIII:Xi1820,LHCbXi16901820} but not all quantum numbers and branching ratios for the different decay channels can be estimated~\cite{PDG}. Both resonances are too light to be accomodated in most quark models~\cite{Capstick:QM1,Blask:QM2} and, particularly for the \XRes state, only the decay in the neutral $\uppi \Xi$ channel has recently been observed by the Belle Collaboration~\cite{BelleXi}, confirming the first experimental evidence in the same channel obtained in the 1970s~\cite{Ross:Xi1620_1Exp,Briefel:Xi1620_2Exp,Bellefon:Xi1620_3Exp}. Due to the lack of experimental data, the nature and the properties of the \XRes are still open for discussion and its theoretical modeling is far away from being settled. Since this state can in principle couple to the \LAK system, the possibility to access the \LAKInt interaction with the femtoscopy technique opens a new road in the study of double-strange resonances.\\
By considering the interaction between a $\Lambda$ and a kaon, the $S=0$ meson--baryon dynamics can be probed, in which, as for the \LAK system, many inelastic channels are present (such as $\uppi$N, \SK). Effective Lagrangians describing the coupled-channel $S=0$ system are mainly anchored to the large database on elastic $\uppi$N scattering~\cite{Chiral_Liu,Chiral_Mai,BonnLK1,BonnLK2}, which leads to a detailed understanding of most of the light-flavor baryonic resonances known today, such as $N^\ast$ and $\Delta$. However, there might also be states which strongly couple to inelastic channels with no net strangeness, such as \LKInt~\cite{Sekihara:LKComp,Bruns:N1535}. Providing experimental constraints on the \LKInt interaction can hence contribute to improve the knowledge of the light hadronic spectrum.

Additional data on the interaction between $\Lambda$ baryons and strange mesons is also important in view of the recent efforts in going beyond the non-interacting picture of hadrons in the statistical approaches applied in heavy-ion collisions (HIC)~\cite{Dashen:SMatrix1,Lo:SMatrix2,Venugopalan:SMatrix3}. The proton-to-pion ratio~\cite{ALICE:ppionratio}, which was not properly reproduced within the basic assumption of thermal models describing the hadronic phase as a non-interacting system, found its explanation in the inclusion of the $\uppi$N scattering parameters in a more sophisticated recent statistical approach~\cite{Andronic:ppuzzle}. Since $\Lambda$ and kaons/antikaons are the most abundant strange hadrons produced in HICs, the interaction between them can be used as an input for these new calculations within the thermal model and help to shed light on the role of strangeness in the hadronization process~\cite{ALICE:StrangeEnh}.\\
Correlations of all the neutral and charged combinations between \lmb and kaons (\LK, \LAK, \LKzero) have been published by the ALICE Collaboration in \PbPb collisions at a center-of-mass energy per nucleon--nucleon collision \twosevensixnn~\cite{LK_PbPbALICE}, and delivered the first scattering parameters on the underlying interaction, being repulsive for \LK and attractive for the remaining pairs. A similar measurement on \LKzero has also been conducted recently by the CMS Collaboration in \PbPb collisions at \fivenn~\cite{CMS:LK0}, in which a different treatment of feed-down contributions is used.

In this Letter, we study the \LKInt and \LAKInt interaction via the measurement of the correlations for the charged combinations \LKpair and \ALKpair in pp collisions at \onethree~\cite{ALICE,ALICEperf}.\\
In order to enhance the number of \LKpair and \ALKpair pairs, the analysis is performed in the high-multiplicity (HM) data sample in which an enhanced yield of strange particles, as \lmb and kaons, is observed~\cite{ALICE:EnhancedStrange}. Note that the correlation functions of \LKPlus (\LKMin) pairs and \ALKMin (\ALKPlus) pairs are added together in order to enhance the statistical significance of the results.
The results are obtained by comparing the experimental data to the modeled correlation using the \Ledn analytical formula, from which scattering parameters and properties of the \XRes are extracted.

\section{Data analysis}\label{sec:DataAnalysis}
The data sample studied in this work was collected by ALICE~\cite{alicecollaboration2022alice} at the LHC in pp collisions at $\sqrt{s} = 13$ TeV during the Run 2 period. All analyzed events passed a HM trigger, based on the measured amplitude in the \VZERO detector system, consisting of two arrays of plastic scintillators located at forward ($2.8<\eta<5.1$) and backward ($-3.7<\eta<-1.7$) pseudorapidities~\cite{VZERO}. The selected events correspond to the inelastic \pp collisions with the top 0.17\% of the measured signal amplitudes, with at least one charged particle in the range $|\eta| < 1$ (referred to as INEL $>$ 0)~\cite{ALICE,ALICEperf}. 
The resulting data sample contains events with an average of 30 produced charged particles in the pseudorapidity interval \etarange{0.5}~\cite{ALICE:pOmega}.Approximately $1.0 \times 10^9$ HM events are selected by adopting
the procedure described in Refs.~\cite{ALICE:pSig0, ALICE:pOmega, ALICE:Source}.

The Monte Carlo simulated data used in this analysis are obtained from the PYTHIA 8.2 event generator~\cite{PYTHIA}.
The transport through the ALICE detector is simulated using GEANT 3~\cite{GEANT3} and the reconstruction follows the dedicated ALICE reconstruction algorithm~\cite{ALICE}.
An additional selection on large charged-particle multiplicities, which mimics the effect of the HM trigger, is applied.

The primary vertex (PV) of the collision is measured using the charged-particle tracks reconstructed from the Inner Tracking System (ITS)~\cite{ALICEITS} and the Time Projection Chamber (TPC)~\cite{ALICETPC}. A maximal displacement of the PV with respect to the nominal interaction point of 10 cm along the beam axis is required in order to ensure a uniform acceptance. Charged particles are identified using information provided by the TPC~\cite{ALICETPC} and the Time-of-Flight (TOF) detector~\cite{ALICETOF}. The ITS, TPC, and TOF detectors, used for charged-particle tracking and identification, cover the full azimuthal angle and the pseudorapidity interval \etarange{0.9}, and are embedded in a uniform magnetic field of 0.5 T along the beam axis.

The information provided by these detectors is used to extract the kinematic and topological 
quantities needed to reconstruct the K (\kbar) and \lmb (\almb) candidates. The selection on these 
variables is varied to evaluate the related systematic uncertainties. In the following, the systematic variations of the selections specifically mentioned in the text are enclosed in parentheses.

The identification of kaons (antikaons) is conducted employing both the TPC and TOF detectors by applying a strict selection on the deviation $n_\sigma$ between the measured quantities ($\textrm{d}E/\textrm{d}x$, time-of-flight) and the signal hypothesis for a kaon, electron, or pion, normalized by the detector resolution $\sigma$. The $n_\sigma$ thresholds are chosen so as to remove possible contamination from electrons and pions to the kaon sample.
The kaon candidates are selected within a transverse momentum range of \pt $\in$ [0.15 (0.1, 0.2), 4.0]~\GeVc and a pseudorapidity range of $|\eta|< 0.8\,(0.75, 0.85)$, to avoid regions of the detector with limited acceptance.
To significantly improve the amount of primary kaons with respect to secondary particles coming from weak decays and particle–detector interactions, a selection criterion on the Distance of Closest Approach (DCA) to the primary vertex is applied, both in the transverse plane ($\mathrm{DCA}_{xy} < 0.1$ cm) and along the direction of the beam ($\mathrm{DCA}_{z} < 0.2$ cm). The purity, referring to the fraction of correctly identified kaon and antikaon candidates, is around 99.5\% and the primary fraction is estimated to be 57.6\%, using the same procedure described in Ref.~\cite{ALICE:pK}.

The kinematic and topological selection criteria related to the reconstruction of \lmb and \almb, as well as the associated systematic uncertainties, are the same as described in Ref.~\cite{ALICE:Source}.
Due to their charge neutrality and their short lifetime, the \lmb  candidates are reconstructed through the weak decay $\mathrm{\lmb \rightarrow p \pi^-}$, which has a branching ratio of BR = (63.9 $\pm$ 0.5)$\%$ and a decay length of $\mathrm{c \tau = (7.89 \pm 0.06)}$ cm~\cite{PDG}. The charge-conjugate decay is used for the \almb  reconstruction. 
The candidates are then identified within a p$\pi^-$ invariant mass window of 
$|M_{\mathrm{p}\pi} - M_{\Lambda}| < 4$~\MeVc$^2$ (corresponding to about 3$\sigma$),
with the nominal mass $M_{\Lambda} = 1116$~\MeVc$^2$~\cite{PDG}. This leads to purities of $P_{\Lambda}$ = $94.2\%$, $P_{\almb}$ = $95.1\%$ for $\Lambda$ and $\almb$, respectively.
A primary fraction of 57.6\% is extracted following the procedure described in Ref.~\cite{pLambda_Dimi}. Secondary contributions from weak decays of neutral and charged $\Xi$ baryons
account for $23.2\%$ of the candidate sample. The remaining 19.2\% are attributed to $\Sigma^0$ particles.

\section{Analysis of the correlation function}\label{sec:cf}
The observable of this analysis is the two-particle correlation function $C(\kstar)$, defined as~\cite{Lisa:2005dd}

\begin{align}\label{eq:defCFexp}
C(\kstar)=\mathcal{N}\times\frac{N_{\rm{same}}(\kstar)}{N_{\rm{mixed}}(\kstar)}, 
\end{align}

where $\kstar = \frac{1}{2}\times|\mathbf{p}^*_1-\mathbf{p}^*_2|$ is the relative momentum of the pair in its rest frame.
Here $N_\mathrm{same}(\kstar)$ is the \kstar distribution of pairs measured in the same event, $N_\mathrm{mixed}(\kstar)$ is the reference distribution of uncorrelated pairs sampled from different (mixed) events. The mixed-event sample is obtained by pairing particles stemming from events with a similar number of charged particles at midrapidity and a close-by primary vertex position along the beam direction, following~\cite{ALICE:pSig0,ALICE:pKpp,ALICE:pOmega}.
The constant $\mathcal{N}$ is a normalization parameter determined by assuming particle pairs with large \kstar to be uncorrelated, which corresponds to a flat $C(\kstar)=1$~\cite{Lisa:2005dd}. This normalization constant $\mathcal{N}$ is evaluated in \kstar $\in$ $\mathrm{[240-340]}$~\MeVc for \LKpair and in the region $\mathrm{[500-800]}$~\MeVc for \ALKpair, where no resonances are present.

\begin{figure}[h!]
    \centering
    \includegraphics[width=0.8\textwidth]{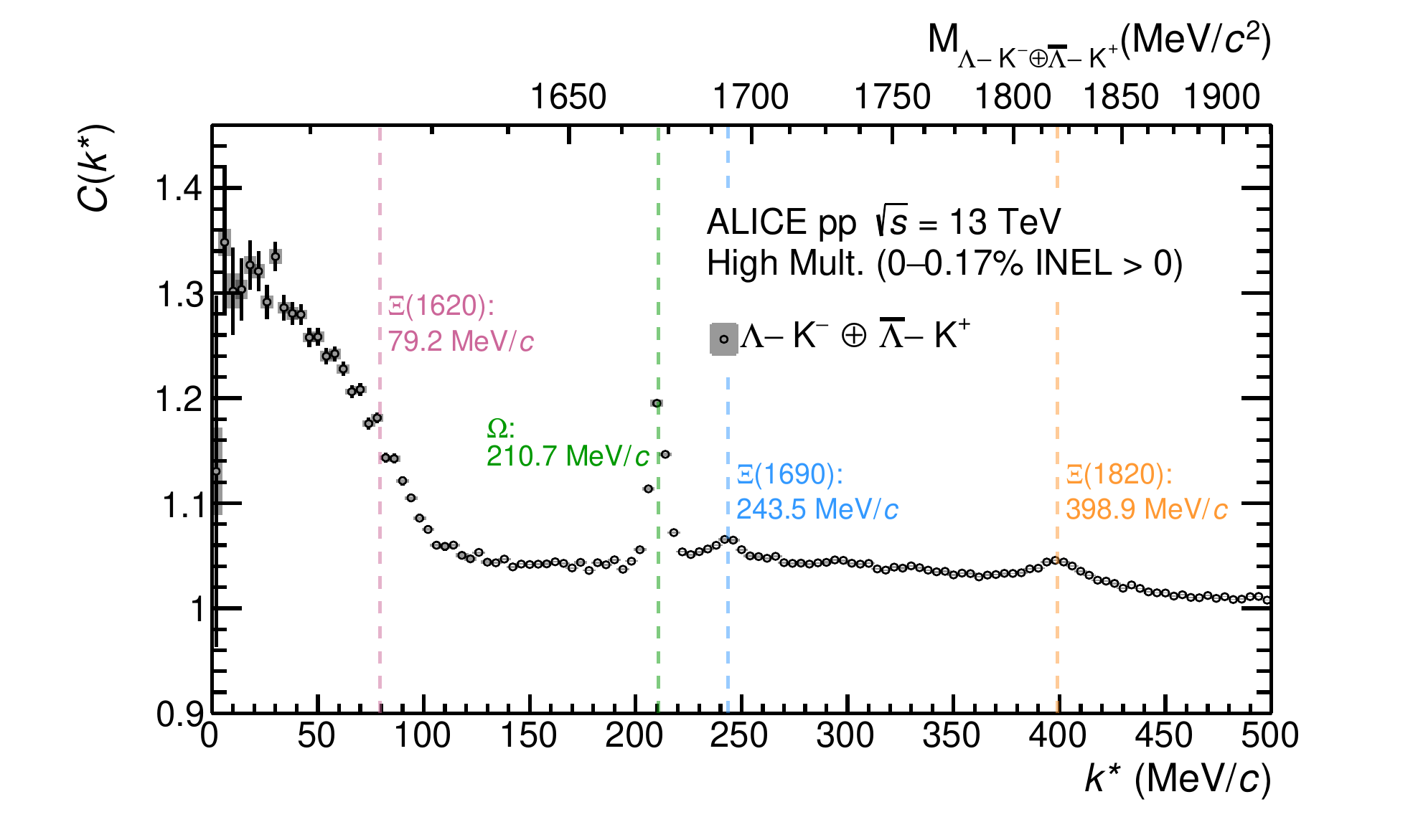}\\
    \includegraphics[width=0.8 \textwidth]{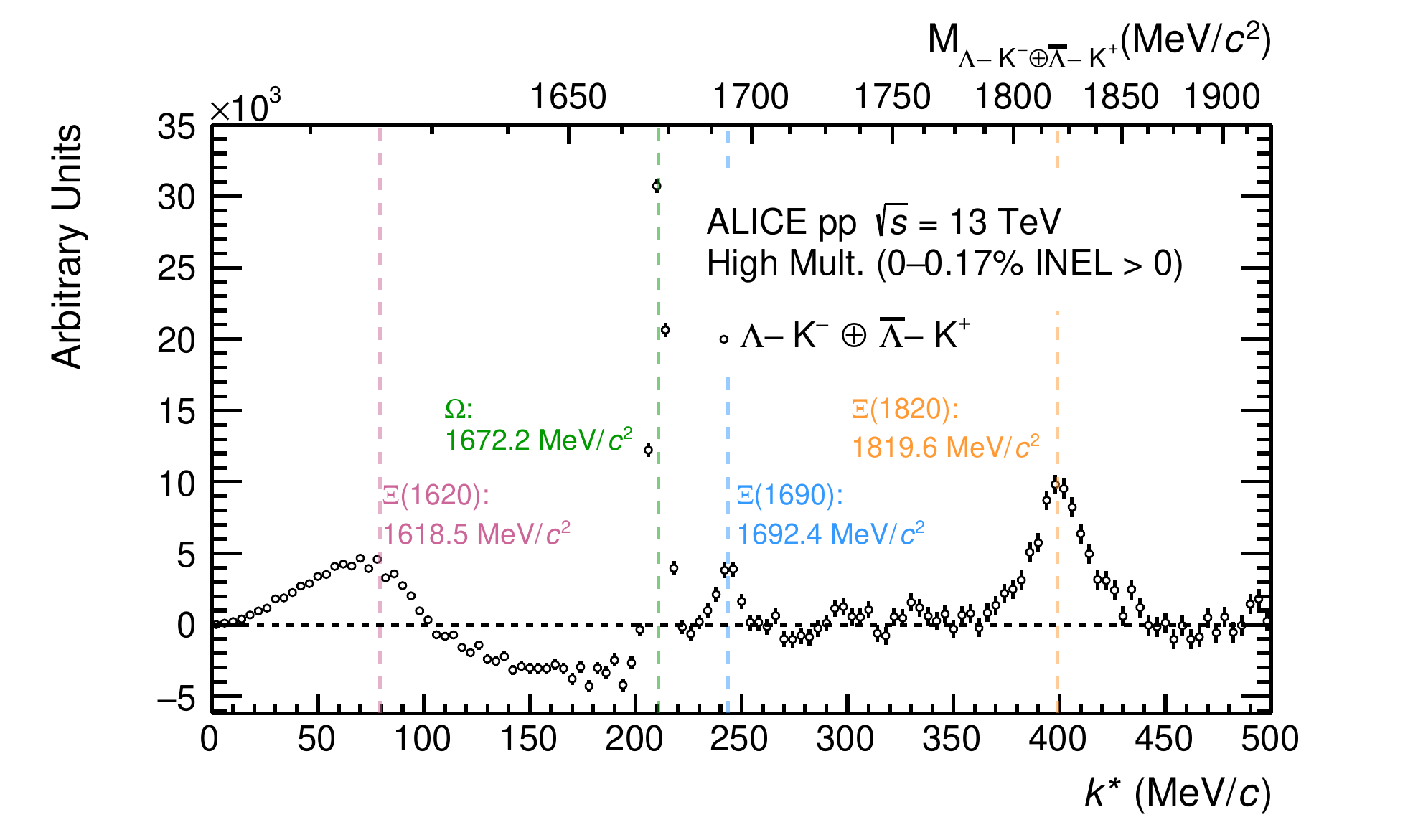}
    \caption{(Color online) Upper: measured correlation function for \LKMin pairs (empty points) with statistical (line) and systematic (gray boxes) uncertainties. Lower: invariant mass spectrum of \LKMin pairs used to build the measured correlation function. Only the statistical uncertainties are shown. The upper \textit{x}-axis indicates the energy at rest $E=\sqrt{(\kstar)^2+m_\lmb ^2}+\sqrt{(\kstar)^2+m_K ^2}$ of the pair written as a function of the relative momentum of the \LKMin pair. The quantity $E$ corresponds to the invariant mass $M$ of the \LKMin pairs. The colored vertical dashed lines indicate the values of the relative momentum \kstar (upper panel) and the value of the energy $E$ at rest of each resonance (lower panel) corresponding to its nominal mass extracted in the final femtoscopic fit.}
    \label{fig:ImCFALK}
\end{figure}

\begin{figure}[t!]
\centering
	\includegraphics[width=0.69\textwidth]{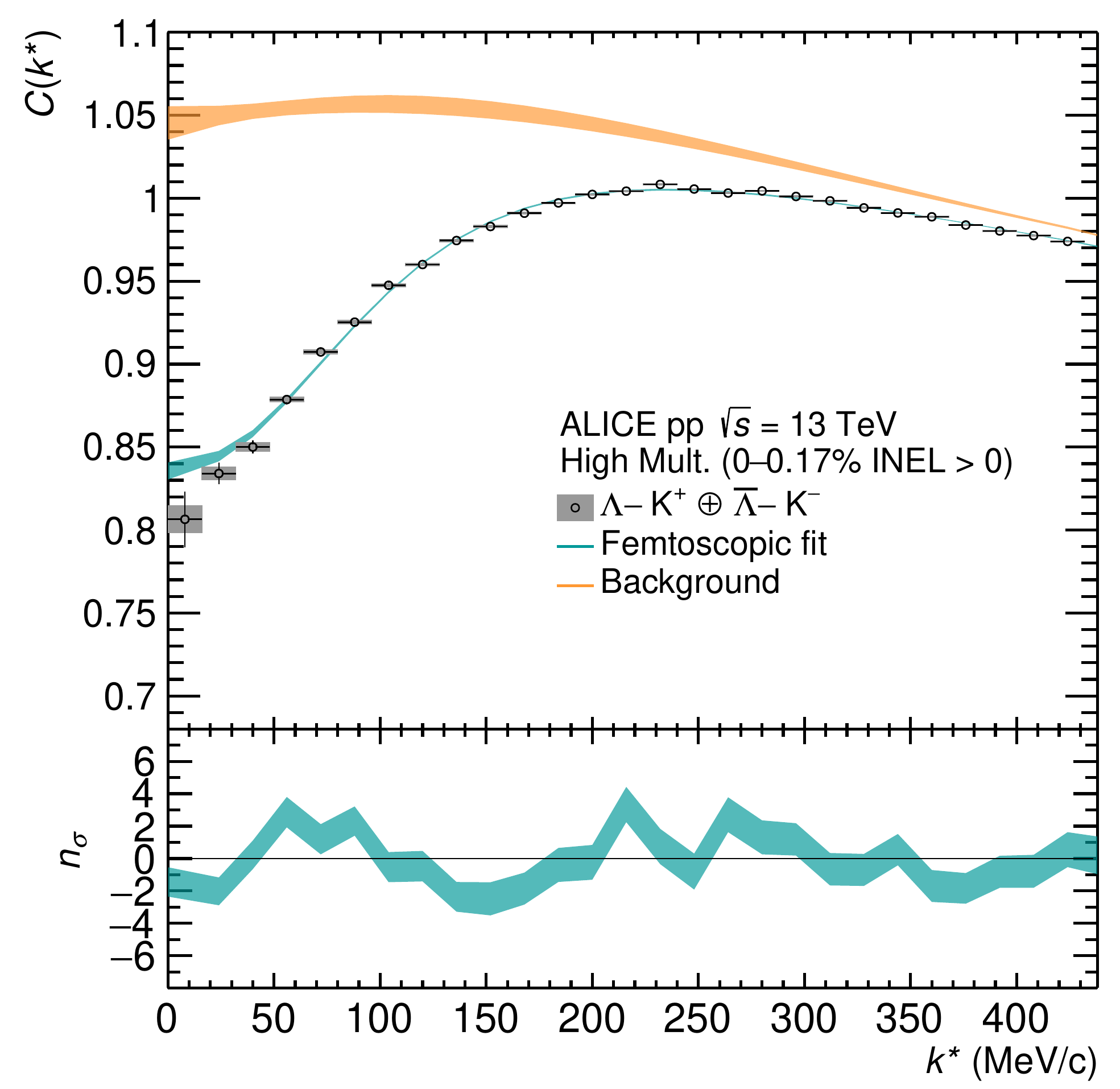}
\caption{(Color online) Measured correlation function of \LKPlus pairs. Statistical (bars) and systematic (boxes) uncertainties are shown separately. 
The light cyan band represents the total fit obtained using Eq.~\ref{eq:totcorrelation} from which the normalization $N_D$, and the scattering parameters (\Rescatt,\Imscatt, and \effran) are extracted. The orange band represents the $C_\mathrm{background}(\kstar)$ contribution, modeled as described in Section~\ref{sec:cf}, and  multiplied by the constant $N_D$. Lower panel: $n_\sigma$ deviation between data and model in terms of numbers of standard deviations.}
\label{fig:Ledn_LKPlus}
\end{figure}

A total of $\num{4.45e6}$ \LKpair and $\num{4.38e6}$  \ALKpair pairs contribute to the correlation signal for $\kstar<200$~\MeVc. For brevity, in the following \LKPlus denotes the combination \LKpair and
\LKMin is used for \ALKpair. The resulting experimental correlation functions are shown in the upper panel of Fig.~\ref{fig:ImCFALK}  and in Fig.~\ref{fig:Ledn_LKPlus}.

The measured correlations are ﬁtted with a correlation function:

\begin{align}\label{eq:totcorrelation}
    C_\mathrm{tot}(\kstar) = & N_D \times C_\mathrm{model}(\kstar) \times C_\mathrm{background} (\kstar),
\end{align}

where $N_D$ is a normalization constant, free to vary in the fit. The default fit range is $0<\kstar<500$~\MeVc. A variation of $\pm10\%$ to the upper limit of the default fit range is applied for evaluating the systematic uncertainties. The term $C_{\mathrm{background}}$ is related to a possible residual background, which can still be present in the femtoscopic correlation. Its modeling is addressed in details later in this section. The term $C_{\rm{model}}(\kstar) = 1 + \sum_{i} \lambda_{i} \times (C_{i}(\kstar) - 1)$ includes the genuine correlation ($i= \mathrm{gen}$), which arises from final state interaction among the two particles of interest, as well as residual contributions involving secondary particles from weak or electromagnetic decays and misidentified ones. Each of these contributions is weighted by the corresponding $\lambda_i$ parameter, evaluated as the
product of the purities and fractions (primary or secondary) of the particles composing the $i$ pair~\cite{ALICE:Run1}. The latter are reported for kaons and \lmb baryons in Sec.~\ref{sec:DataAnalysis}.

The genuine contribution for \LKPlus and \LKMin pairs amounts to $\lambda_{\mathrm{gen}}=51\%$; the residual correlations between kaons (antikaons) and \lmb (\almb) from the decay of $\Xi$ ($\overline{\Xi}$) contribute each with a weight of $\lambda_{\Lambda_{\Xi}K}=10\%$. The correlations for the charged combinations (e.g~$\Xi^{\pm}\mbox{--}\mathrm{K^{\pm}}$) are modeled with the CATS framework~\cite{CATS} assuming Coulomb-only interaction. The presence of a residual strong interaction between $\Xi$ and kaons is neglected in this analysis since currently no experimental data are available and the corresponding theoretical predictions are hence not validated yet~\cite{Wang:XiK,Nogueira-Santos:XiK}. Similarly, residual correlations involving $\Xi^0$ and $\Sigma^0$ are considered to be flat due to the absence of Coulomb interaction.
Such residual contributions, along with correlations involving misidentified particles, amount to $\lambda_{\mathrm{flat}}=39\%$ of the measured signal.
The systematic uncertainties related to the $\lambda_i$ parameters are estimated based on the purities obtained for each varied set of kinematic and topological cuts, as well as by varying the values of secondary fractions by $\pm10\%$. In addition to the feed-down contributions, a correction for the finite experimental momentum resolution is taken into account for a direct comparison with data~\cite{ALICE:Run1}.

The last factor in Eq.~\ref{eq:totcorrelation}, $C_{\mathrm{background}}$, accounts for the non-femtoscopic background visible in both measured correlations~\cite{ALICE:Run1}. In particular, the \LKPlus data are affected by the presence of the so-called mini-jet contributions, typically associated to the initial hard processes occurring at the parton level during the collision~\cite{ALICE:minijet}. This type of background has already been observed in several meson--meson\cite{Abelev:2012151, Abelev:2012sq,PhysRevD.84.112004, PhysRevC.91.034906}, meson--baryon~\cite{ALICE:pphi,ALICE:pKpp}, and baryon--antibaryon femtoscopic analyses~\cite{Vale_BBar} and it is particularly enhanced when the net-charges, such as baryon number, electric charge and strangeness, are zero for the pair at hand. The mini-jet term included in the $C_{\rm{background}}(\kstar)$ for \LKPlus is modeled using Monte Carlo simulated data and following the same procedure adopted in Ref.~\cite{Vale_BBar}. 
A polynomial of second order is added as baseline to the mini-jet part of the background to take into account energy--momentum conservation effects developing at large $\kstar$\cite{ALICE:Run1}, which lead to an enhancement of the correlation function in this momentum region. 
The coefficients of the polynomial are fixed by a prefit of $C_{\mathrm{background}}(\kstar)$ to the \LKPlus data in the region of $400< \kstar< 2000$~\MeVc. A variation of $\pm10\%$ in the lower and upper limit of this range range is included to estimate the systematic uncertainty related to the total background. In the case of \LKMin, the mini-jet background is much less pronounced since the the net-strangeness of the pair is not zero. The measured \LKMin correlation function outside the femtoscopic region of $\kstar > 200$~\MeVc is well reproduced by Monte Carlo simulated data and no additional baselines are required to describe the large \kstar region. Hence, the total $C_{\mathrm{background}}(\kstar)$ in the \LKMin case is modeled using only the simulated correlation, parametrized by a third-order polynomial constrained to
be flat at $\kstar\rightarrow 0$ since no signal is expected to arise from the background~\cite{ALICE:pL,ALICE:LXi} at low \kstar. The coefficients of the polynomial are fixed by fitting the Monte Carlo simulated data in the range \kstar $\in$ $[0,600]$~\MeVc, with systematic variations of the upper limit of $\pm10\%$.

Besides the background, which shifts the data upwards with respect to unity in the region of relative momenta above 200~\MeVc, there are several structures present in the measured \LKMin correlation (upper panel in Fig.~\ref{fig:ImCFALK}), related to resonances decaying into \LKMin pairs.
In the lower panel of Fig.~\ref{fig:ImCFALK}, the invariant mass of \LKMin pairs, expressed in \kstar, is shown in order to better visualize the location of these three resonances. The \LKMin invariant mass spectrum in \kstar is obtained using the same- and mixed-event distributions in Eq.~\ref{eq:defCFexp} and by following the same approach employed in resonance analyses~\cite{ALICE:res1}.
The uncorrelated mixed-event distribution, normalized to the signal outside the resonance region (\kstar $\in$ $[500,800]$~\MeVc), is subtracted from the same-event \LKMin signal. After this subtraction, a residual background is still present, corresponding to the $C_{\mathrm{background}}(\kstar)$ contribution in the measured correlation function shown above. This remaining background in the invariant mass distribution can be modeled as well using Monte Carlo data, fitted with a fourth-order polynomial, and it can be finally subtracted from the measured spectrum in order to investigate in more details the \LKMin dynamics at low \ks. 
The peak appearing at $\kstar\approx 211$~\MeVc (green dashed line) corresponds to the $\Omega$ baryon decaying weakly into \LKMinInt with a branching ratio of $(67.8\pm 0.7)\%$~\cite{PDG}. The $\Xi(1690)$ and $\Xi(1820)$ resonances can be associated to the second and third peaks in the correlation. The negative counts in the region $100 < \kstar < 200$ \MeVc of the \LKMin invariant mass are due to the presence of a residual non-resonant interaction, which will be discussed in detail in Sec.~\ref{sec:Results}. \\
Currently the branching ratios of the strong decays of these states into \LKMinInt are not precisely known~\cite{PDG}. In the upper \textit{x}-axis of the bottom panel of Fig.~\ref{fig:ImCFALK}, the corresponding mass of the resonance obtained from the kinematic relation $E=\sqrt{(\kstar)^2+m_\lmb ^2}+\sqrt{(\kstar)^2+m_K ^2}$, is shown. In order to properly model the background outside the femtoscopic range, these three resonances must be taken into account.
The total background correlation for \LKMin can hence be written as

\begin{align}\label{ALK_Cback}
    C_{\mathrm{background}}^{\Lambda \mathrm{K}^{-}}(\kstar) = \alpha_\mathrm{pol3} (1+ b \kstar {^2} +c  \kstar{^3}) + \alpha_\Omega  f_\mathrm{G} (M_\Omega,\sigma_\Omega) + \sum_i \alpha_i f_\mathrm{BW}(M_i,\Gamma_i),
\end{align}

in which a Gaussian distribution $f_\mathrm{G}$ is used for the $\Omega$ baryon decaying weakly to \LKMin,  and a typical Breit-Wigner one $f_\mathrm{BW}$ for the two excited $\Xi$ resonances, having broader widths due to the strong decay to \LKMin
. In order to help the convergence of the final femtoscopic fit, a fit of the total $C_{\mathrm{background}}(\kstar)$ correlation to the data is performed in the \kstar region of $190-600$~\MeVc to estimate the weights $\alpha_\Omega$, $\alpha_i$ as well as the masses and widths of the resonances. A change of $\pm10\%$ in the upper limit of the prefit range is included in the evaluation of the final systematic uncertainties. These parameters are then kept free in the final femtoscopic fit of $C_{\mathrm{tot}}(\kstar)$ to the data and the values obtained for the masses and widths are found to be compatible with the available PDG values~\cite{PDG} and recent measurements~\cite{LHCbXi16901820,BESIII:Xi1820}. The orange band in Figs.~\ref{fig:Ledn_LKPlus} and~\ref{fig:Ledn_ALKMin} shows the total $C_{\mathrm{background}}(\kstar)$ correlation function extracted in the final femtoscopic fit, multiplied by the normalization factor $N_D$, for \LKPlus and \LKMin pairs, respectively.

The last ingredient needed to model the data is the strong interaction of the \LKPlus and \LKMin pairs entering in the $C_{\rm{model}}(\kstar)$ in Eq.~\ref{eq:totcorrelation} via the genuine correlation function $C_\mathrm{gen}(\kstar)$. This is modeled for both pairs using the \Ledn analytical formula~\cite{Lednicky:1981su}, following the approach used in Ref.~\cite{LK_PbPbALICE},

\begin{align}\label{eq:LednFormula}
    C(\kstar)_\mathrm{LL} = 1 + \Bigg[\frac{1}{2} \Bigg|\frac{f(\kstar)}{R}\Bigg|^2 \Bigg( 1 - \frac{\effran}{2 \sqrt{\pi}R} \Bigg) + \frac{2 \RescattAmpl(\kstar) }{\sqrt{\pi} R} F_1(2\kstar R) - \frac{\ImscattAmpl(\kstar)}{R}F_2(2\kstar R)\Bigg]. 
\end{align}
The scattering amplitude $f(\kstar)$ is the quantity embedding the scattering parameters and providing information on the underlying interaction. Typically, $f(\kstar)$ is expressed via the effective-range expansion (ERE) $f(\kstar)=\left( \frac{1}{f_0}+\frac{1}{2}d_0 k ^{\ast ^2} -i\kstar \right)^{-1}$, in which $f_0$ is the scattering length and $d_0$ is the effective range. The parameter $R$ is the size of the emitting source with a Gaussian profile. In this work it was fixed using the core-resonance model taken from Ref.~\cite{ALICE:Source}, already employed in several previous femtoscopic analyses performed in small colliding systems as \pp collisions and anchored to \pprot correlations. The core radius for \LKPlus and \LKMin pairs is $r_\mathrm{core}(\left<m_\mathrm{T}\right>=1.35 \,\,\mathrm{Ge\kern-.2emV/}c^2)=1.11\pm0.04~$ fm. In order to use the core-resonance total source in Eq.~\ref{eq:LednFormula}, this must be parametrized with a Gaussian distribution. The presence of long-lived strong resonances feeding to \lmb and kaons introduces a significant exponential tail for large \rstar, which cannot be described with a single Gaussian~\cite{ALICE:pL,ALICE:pOmega,ALICE:LXi,ALICE:pphi,ALICE:pKCoupled}. The total source is hence modeled with a weighted sum of two Gaussians, leading to an effective emitting source $S_\mathrm{eff}(\rstar)=\lambda_S[\omega_S S_1(\rstar)+(1-\omega_S)S_2(\rstar)]$, in which $r_{1}=1.202^{+0.043} _{-0.042}$ fm , $r_{2}=2.330^{+0.050} _{-0.045}$ fm, $\lambda_S=0.9806^{+0.0006} _{-0.0008}$, and $\omega_S=0.7993^{+0.0037} _{-0.0027}$. As systematic variation of the source function, these values are varied within the uncertainties. Due to the additive property of correlation functions, the final genuine correlation is then taken as the sum of two correlations evaluated with the two properly weighted Gaussian sources. To preserve the correct normalization of the emitting source and the unitarity of the $\lambda$ parameters~\cite{ALICE:Run1} in $C_\mathrm{model}(\kstar)$, a (1$-\lambda_S$) contribution is added.

The understanding of the \LKMinInt interaction, particularly in the low \kstar region, is strictly connected to the \XRes state. In principle, since \XRes shares the same quantum numbers as the \LKMin pair,  the two systems can couple strongly. The Belle collaboration recently published the observation of the \XRes state in the $\Xi\uppi$ decay channel ($E_\mathrm{thr.1}=m_\uppi +m_\Xi=1461.3$ \MeVmass)~\cite{BelleXi}. The reported mass and widths in Ref.~\cite{BelleXi} are $M_{\XRes}=1610.4\pm 6.0$ \MeVmass, $\Gamma_{\XRes}=60.0\pm4.8$ MeV, which indicates that the decay of \XRes into \LKMinInt ($E_\mathrm{thr.2}=m_\mathrm{K^-} +m_\Lambda=1609.4$ \MeVmass) is kinematically allowed.
 No experimental evidence of this decay channel has been observed so far. The presented work provides quantitative evidence of this process.
 
 The \XRes state can be clearly seen in the peak at $\kstar \approx 80$~\MeVc in the lower panel of Fig.~\ref{fig:ImCFALK}. Hence, to model the \LKMinInt interaction at low \kstar, the \XRes must be taken into account in the \Ledn approach. Similar scenarios, with resonances contributing to the signal in the low \kstar region, were observed in $\mathrm{K^0 _S-K^{\pm}}$ correlations measured in \pp and \PbPb collisions, in which the interaction mainly goes through the formation of the $a_0$ resonance. A way to properly include such a resonant interaction is to write the scattering amplitude in Eq.~\ref{eq:LednFormula} in terms of the probability distribution describing the state. Due to the vicinity of the \LKMinInt decay-channel threshold, the \XRes resonance must be described with a Flatté-like distribution~\cite{Flatte:1976xu} such as the Sill distribution used in Ref.~\cite{GiacosaSill}. The corresponding scattering amplitude can be written as

\begin{align}\label{eq:scattAmplSill}
    f(\kstar) = & \dfrac{-2\Tilde{\Gamma}_{\mathrm{\Lambda K^{-}}}}{E^2-M^2+i\Tilde{\Gamma}_{\mathrm{\Xi\uppi}}\sqrt{E^2-E_{\mathrm{thr.\mathrm{\Xi\uppi}}} ^2}+i\Tilde{\Gamma}_{\mathrm{\Lambda K^{-}}}\sqrt{E^2-E_{\mathrm{thr.\mathrm{\Lambda K^-}}} ^2}}
\end{align}

in which $M$ is the mass of the \XRes state, $\Tilde{\Gamma}_{i=\mathrm{\Xi\uppi},\mathrm{\Lambda K^-}}$ are the effective partial widths as defined in Ref.~\cite{GiacosaSill}, and $E_{\mathrm{thr}.i=\mathrm{\Xi\uppi},\mathrm{\Lambda K^-}}$ are the threshold energies for the two channels, as defined above.

Besides the interaction between \lmb and antikaons through the \XRes state, a non-resonant strong interaction is present in the measured correlation function, which can be explicitly seen in the lower panel of Fig.~\ref{fig:ImCFALK} for $100 < \kstar < 200$~\MeVc. In this \kstar region the data, corrected by the background contribution as described above, go below zero indicating a depletion in the measured \LKMin pairs arising from the underlying non-resonant component of the interaction. Since there are no theoretical approaches available at the moment in which the \LKMinInt interaction is composed of a resonant part, through the \XRes state above the \LKMinInt threshold, and a non-resonant one, an effective modeling of these two contributions will be adopted employing the \Ledn formula. The non-resonant $C_\mathrm{LL} ^\mathrm{ non\text{-}res}(\kstar)$ and resonant $C_\mathrm{LL} ^\mathrm{res}(\kstar)$ correlations are modeled using Eq.~\ref{eq:LednFormula}: for $C_\mathrm{LL} ^\mathrm{ non\text{-}res}(\kstar)$ an ERE scattering amplitude is assumed, while for $C_\mathrm{LL} ^\mathrm{res}(\kstar)$ a Sill amplitude is employed, according to Eq.~\ref{eq:scattAmplSill}.
Taking both interactions into account, the final genuine correlation for \LKMin is composed of a weighted sum of a correlation including the resonant process and another one responsible for the non-resonant part

\begin{align} \label{eq:totgenALK}
    C_\mathrm{gen}(\kstar) = \omega C_\mathrm{LL} ^\mathrm{ non\text{-}res}(\kstar)+(1-\omega) C_\mathrm{LL} ^\mathrm{res}(\kstar).
\end{align}

The remaining free parameters to be extracted in the final femtoscopic fit of $C_\mathrm{tot}(\kstar)$ to the data are the weight $\omega$ for non-resonant scattering parameters $(
\Rescatt,\Imscatt,\effran)$, the mass $M$, the partial widths $\Tilde{\Gamma}_{i=1,2}$ of the \XRes state, and the masses and widths of the $\Omega$, \XResNovanta, and \XResVenti.

\section{Results}\label{sec:Results}
\begin{figure}[t!]
\centering
	\includegraphics[width=0.69\textwidth]{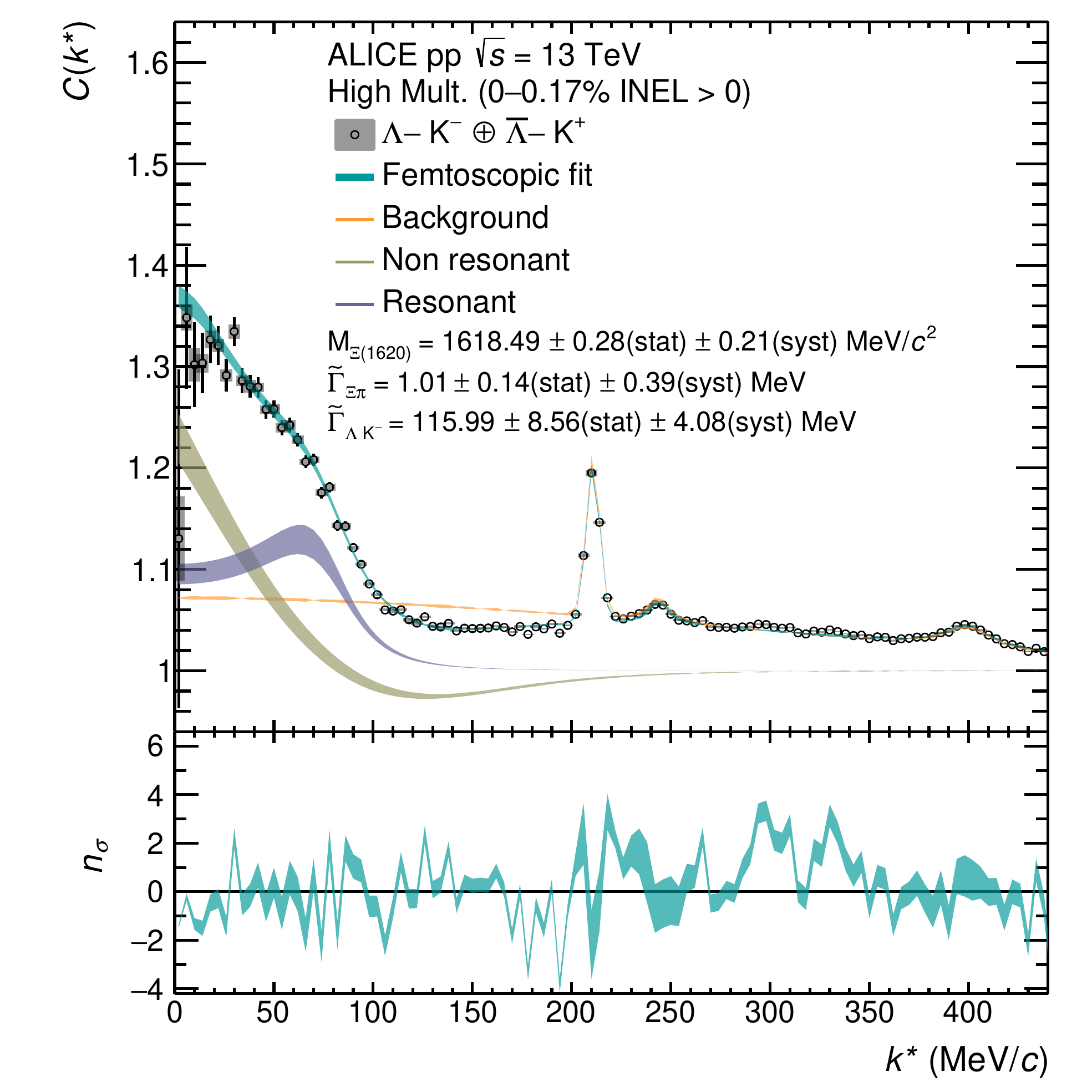}
\caption{(Color online) Measured correlation function of \LKMin pairs. Statistical (bars) and systematic (boxes)
uncertainties are shown separately. 
The light cyan band represents the total fit obtained using Eq.~\ref{eq:totcorrelation} from which the normalization $N_D$, the non-resonant scattering parameters (\Rescatt,\Imscatt and \effran) and the properties of the \XRes state are extracted. The violet band represents the $C_\mathrm{LL} ^\mathrm{res}(\kstar)$ correlation multiplied by the corresponding weight $(1-\omega)$, while the olive green band is the $\omega C_\mathrm{LL} ^\mathrm{ non-res}(\kstar)$. The orange band represents the $C_\mathrm{background}(\kstar)$ modeled using the Monte Carlo simulations multiplied by the constant $N_D$. Lower panel: $n_\sigma$ deviation between data and model in terms of numbers of standard deviations.}
\label{fig:Ledn_ALKMin}
\end{figure}

The results for \LKPlus and \LKMin systems are shown in Figs.~\ref{fig:Ledn_LKPlus} and~\ref{fig:Ledn_ALKMin}, respectively. The lower panels in each plot show the deviation between data and model in terms of number of standard deviations $n_\sigma$.
The width of the band represents the total uncertainty of the fit, including the statistical and the systematic uncertainties. The gray boxes correspond to the systematic uncertainties of the data and they maximally amount to 3$\%$--4$\%$ in the lowest \kstar interval for each pair.
The measured \LKPlus correlation function, shown in Fig.~\ref{fig:Ledn_LKPlus}, is below unity at low \kstar, indicating a repulsive strong interaction between \lmb and kaons, in agreement with the femtoscopic results obtained in \PbPb collisions ~\cite{LK_PbPbALICE}. The behavior of the data is well reproduced by the fit using Eq.~\ref{eq:totcorrelation} with an average reduced $\chi^2/$NDF of 3.9 estimated in the default fit range.

In Fig.~\ref{fig:Ledn_ALKMin}, the results for the \LKMin system are presented. The light cyan band represents the total correlation function (Eq.~\ref{eq:totcorrelation}) with the genuine interaction modeled, including a non-resonant and a resonant contribution through the formation of the \XRes.  The fit well describes the data and the reduced $\chi^2/$NDF, evaluated within the fit range, is 2.9. The obtained weight $\omega$ in Eq.~\ref{eq:totgenALK} is found to be $0.950\pm 0.005\mathrm{(stat.)}\pm 0.006\mathrm{(syst.)}$, indicating that a dominant contribution from the non-resonant interaction is needed to reproduce the data. However, the approach taking into account both contributions, which is used in this work, should be considered as a phenomenological approach and more theoretical investigations are needed in order to provide a better description of the interplay between resonant and non-resonant processes. The two additional bands reported in Fig.~\ref{fig:Ledn_ALKMin} correspond to the weighted correlation functions obtained from the fit, which represent the resonant (violet) and non-resonant (olive) \LKMinInt interaction, respectively.

The \LKMin pairs interacting through the resonance lead to a rather flat 
correlation profile at low \kstar, which peaks at the mass of the observed \XRes ($\kstar \approx 80$ \MeVc) and then quickly reaches unity. The extracted values of mass and partial effective widths for the \XRes from the femtoscopic fit are also reported in Fig.~\ref{fig:Ledn_ALKMin}. The mass $M_{\XRes}=1618.49\pm0.28\mathrm{(stat.)}\pm0.21\mathrm{(syst.)}$ \MeVmass obtained from the fit, as stated in Sec.~\ref{sec:cf}, is in agreement with the Belle measurement~\cite{BelleXi}. The numerical values of $\Tilde{\Gamma}_{\mathrm{\Xi\uppi}}=1.01\pm0.14\mathrm{(stat.)}\pm0.39\mathrm{(syst.)}$ MeV and  $\Tilde{\Gamma}_{\mathrm{\Lambda K^-}}=115.99\pm8.56\mathrm{(stat.)}\pm4.08\mathrm{(syst.)}$ MeV, obtained from the fit using a Flatté-like distribution, should not be taken literally since mainly the ratio between the two coupling constants to these channels is constrained by near-threshold data~\cite{Baru:2004xg}. Nevertheless, the large $\Tilde{\Gamma}_{\mathrm{\Lambda K^-}}$  indicates a strong coupling of the \XRes state to the \LKMinInt channel. To provide a qualitative comparison between the results obtained in this work and the total width reported by Belle, the determination of the poles for the Sill scattering amplitude in Eq.~\ref{eq:scattAmplSill} is performed. By inserting the values of the extracted mass and widths in the denominator and searching for its zeros, the pole position for the \XRes resonance
corresponds to a state with a mass $M=1616.34^{+0.01} _{-0.05}$ \MeVmass and a total width of $\Gamma=12.00\pm1.24$ MeV. The total uncertainty reported is propagated in the calculation from the quadratic sum of the statistical and systematic error on the mass and widths obtained from the femtoscopic fit. The value of the mass obtained from the pole is compatible with the results reported by Belle, while the width $\Gamma$ is smaller. This discrepancy can arise, as mentioned above, from the presence of a close-by threshold and the typical interpretation of the mass and in particular of the width obtained directly from the energy of the pole might not hold for a near-threshold resonance, as the \XRes in this case.

\begin{figure}[t!]
\centering
	\includegraphics[width=0.79\textwidth]{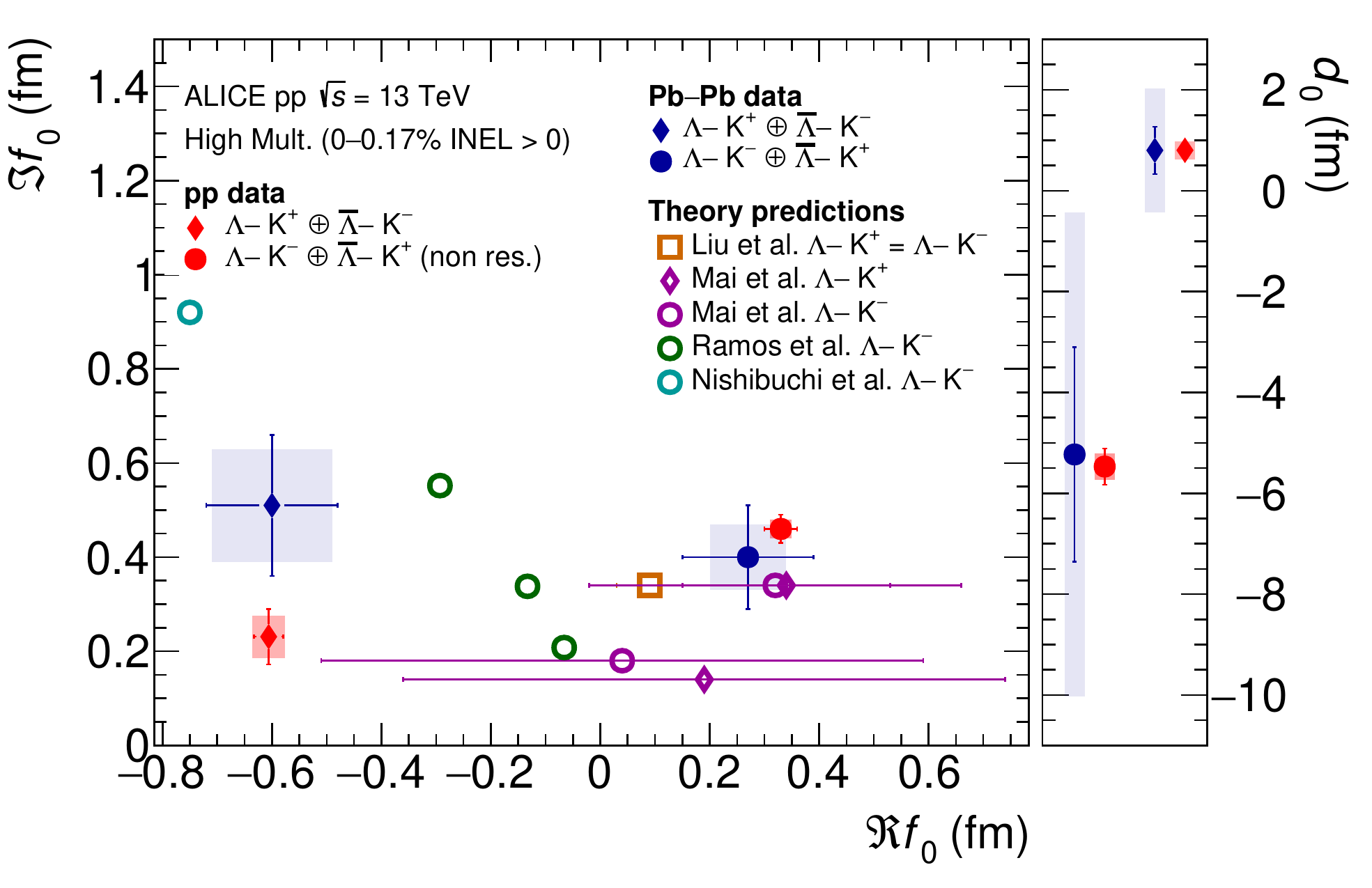}
\caption{(Color online) Left: extracted \Rescatt and \Imscatt for the \LKPlusInt (red diamonds) and \LKMinInt (red dots) interaction in pp collisions, compared to \PbPb results (blue)~\cite{LK_PbPbALICE} and available models (orange~\cite{Chiral_Liu}, magenta~\cite{Chiral_Mai}, green~\cite{ALK_Ramos}, and light cyan~\cite{ALK_Hyodo,ALK_HyodoPaper}). For ~\cite{ALK_Ramos}, the scattering parameters obtained for different sets of input parameters are reported. Statistical (bars) and systematic (boxes) uncertainties are shown. Right: extracted effective range \effran obtained in this work (red) and in \PbPb collisions (blue)~\cite{LK_PbPbALICE}.}
\label{fig:scat_pars_LKPlus}
\end{figure}

The ERE of the scattering amplitude $f(\kstar)$ in the \Ledn formula allows the scattering parameters \Rescatt, \Imscatt, and \effran to be extracted. The results for \LKPlusInt (red diamonds) and for the non-resonant \LKMinInt interaction (red circles) are shown in Fig.~\ref{fig:scat_pars_LKPlus} with statistical (bars) and systematic (shaded areas) uncertainties, and summarized in Table~\ref{tab:scattpars_LK}. The left panel of Fig.~\ref{fig:scat_pars_LKPlus} shows the real part of the scattering length \Rescatt (\textit{x}-axis) as a function of the imaginary part \Imscatt (\textit{y}-axis) obtained in this work and its comparison to  the measurements in \PbPb collisions (blue markers)~\cite{LK_PbPbALICE}. The available theoretical predictions for \Rescatt and \Imscatt, also presented in Fig.~\ref{fig:scat_pars_LKPlus}, are based on unitarized chiral perturbation theory at leading order (LO) (green open circles~\cite{ALK_Ramos}, light-cyan open circles~\cite{ALK_Hyodo,ALK_HyodoPaper}) and on the standard chiral perturbation theory at next-to-leading order (NLO) (orange open squares~\cite{Chiral_Liu}, magenta open markers~\cite{Chiral_Mai}). The output of these chiral calculations strongly depends on the so-called low-energy and subtraction constants, parameters of the model which need as input the experimental data to be fixed. In particular, the results obtained in Ref.~\cite{Chiral_Liu,Chiral_Mai} arise from a full treatment of the $SU(3)$ flavor meson--baryon interaction and are hence anchored to large $|S|=0$ pion--nucleon database. Moreover, to reduce the number of input parameters, isospin symmetry is assumed, and then the crossing symmetry leads to almost identical scattering parameters for the \LKPlusInt and \LKMinInt interaction. In the unitarized framework of~\cite{ALK_Ramos,ALK_Hyodo,ALK_HyodoPaper}, the \XRes is dynamically generated, meaning that the state is not introduced in the Lagrangian investigated at LO, but it appears in the scattering amplitude due to the meson--baryon interaction dynamics. The results in Ref.~\cite{ALK_Ramos} were published before the Belle measurement on \XRes~\cite{BelleXi}, hence the model parameters  were constrained mainly by symmetry assumptions and by experimental data on the antikaon--nucleon interaction. The scattering amplitude obtained within this work has been studied in Ref.~\cite{ALK_Hyodo,ALK_HyodoPaper}, allowing for variations in the subtraction constants of the model in order to reproduce the \XRes properties measured by Belle. An agreement with the measured \XRes mass and width is achieved  only with very large values of these constants, in contrast with the typical trend seen in similar works. Such discrepancy might arise from the inclusion of only LO contributions in the interaction and it can be investigated in the future with an updated version of the chiral potentials in Ref.~\cite{ALK_Feijoo} at NLO that recently became available. The studies in Ref.~\cite{ALK_Ramos,ALK_Hyodo,ALK_HyodoPaper} show that the \LKMinInt interaction is crucial in the understanding of the \XRes state and it must be properly taken into account in the dynamics.

The negative value of \Rescatt obtained in this work for \LKPlus pairs is compatible with the \PbPb results~\cite{LK_PbPbALICE}, confirming the presence of a repulsive interaction. This is in tension with the chiral calculations~\cite{Chiral_Liu,Chiral_Mai}, indicating an overall attraction. 
The extracted \LKPlusInt \Imscatt is in agreement within 1$\sigma$ with the value obtained from the \PbPb measurements and it is in line with the available predictions. The non-negligible value (roughly $1/3$ of the \Rescatt), reported in Table~\ref{tab:scattpars_LK}, indicates the presence of inelastic channels in the measured interaction.
The \effran is shown in the right panel of Fig.~\ref{fig:scat_pars_LKPlus} and it can be seen that, for both \LKPlus and \LKMin systems, the values extracted in \pp and \PbPb colliding systems are in agreement within uncertainties.

The \LKMin pairs undergoing the non-resonant interaction in Fig.~\ref{fig:Ledn_ALKMin} (olive green bands) show an attractive interaction with a depletion in the region $100 < \kstar < 200$ \MeVc (as seen in the lower panel of Fig.~\ref{fig:ImCFALK}) given by a non-negligible \Imscatt.
The \Rescatt lies in the positive side of the \textit{x}-axis, hence indicating an attractive \LKMinInt strong interaction as observed in \PbPb collisions. Similarly, the \Imscatt is compatible within uncertainties with \PbPb results. The agreement between the non-resonant \LKMinInt scattering parameters obtained in this work and the ones obtained in \PbPb collisions is expected since in large colliding systems the effect of resonances, such as the \XRes, on the measured correlation function is suppressed. Hence, the femtoscopic signal will be driven by the non-resonant contribution. The theoretical models currently available on the \LKMinInt interaction are far from converging to a common description of this system, as can be seen in the predicted values of scattering parameters in Fig.~\ref{fig:scat_pars_LKPlus}. As mentioned above, the tension mainly arises from the extremely scarce amount of data available on this system. The work presented in this Letter will significantly improve the knowledge on the interactions between \lmb and antikaons and shed light on the role played by the \XRes state, observed for the first time in this decay channel.

\begin{table}[h!]
\centering
\caption{Extracted scattering parameters for \LKPlusInt interaction and for the non-resonant part of \LKMinInt interaction in \pp collisions. Statistical and systematic uncertainties are reported.}
\label{tab:scattpars_LK}
{\begin{tabular}{ccc}
\hline
Pair & \LKPlus & \LKMin \\ \hline
\rule{0pt}{4ex} \Rescatt (fm) & $-0.61\pm0.03(\mathrm{stat})\pm0.03(\mathrm{syst})$ & 
$0.33\pm0.03(\mathrm{stat})\pm0.02(\mathrm{syst})$ \\
\rule{0pt}{4ex}  \Imscatt (fm) &
$0.23\pm0.06(\mathrm{stat})\pm0.04(\mathrm{syst})$ &
$0.46\pm0.03(\mathrm{stat})\pm0.02(\mathrm{syst})$ \\
\rule{0pt}{4ex} \effran (fm) &
$0.80\pm0.19(\mathrm{stat})\pm0.18(\mathrm{syst})$ &
$-5.47\pm0.36(\mathrm{stat})\pm0.26(\mathrm{syst})$ \\
\hline 
\end{tabular}}
\end{table}

\section{Summary}
\label{sec:summary}
The two-particle correlation technique is used to access the strong interaction between \lmb hyperons and charged kaons. This is achieved by measuring the \LKPlus and \LKMin correlation functions in \pp collisions, down to zero momenta. The results presented in this work provide the most precise data on these interactions. The \LKPlusInt interaction is found to be repulsive, with a non-negligible \Imscatt, indicating that inelastic channels are present. The measured \LKMin correlation function shows a signal above unity at low \kstar, pointing to an overall attractive interaction, as well as several resonances at different \kstar values above 200 \MeVc. The masses and widths of these states, extracted from the fit to the correlation function, are compatible with the $\Omega$ baryon and with two excited $\Xi$ states: the \XResNovanta and the \XResVenti. The invariant mass spectrum of the \LKMin pairs, obtained from the same- and mixed-event distributions entering the measured correlation function shows an additional peak at $\kstar\approx 80$ \MeVc. This structure corresponds to the \XRes state, expected to couple to the \LKMin system. The \XRes is observed so far only in the $\Xi\uppi$ channel~\cite{BelleXi,Ross:Xi1620_1Exp,Briefel:Xi1620_2Exp}, hence these data represent the first experimental evidence of the decay of the \XRes into \LKMin pairs.
The measurements performed in this work show that the \XRes plays an important role at the level of the strong \LKMinInt interaction. To reproduce the measured \LKMin correlation two contributions must be taken into account in the genuine interaction: a resonant term, in which \LKMin pairs interact through the formation of \XRes and modeled via a Sill distribution~\cite{GiacosaSill}, and a residual non-resonant part. Both contributions are modeled using the \Ledn analytical formula with different scattering amplitudes $f(\kstar)$.
The extracted mass and partial widths for the \XRes state are expressed in terms of poles of the Sill scattering amplitude. The mass is found to be consistent with the recent Belle measurements in the $\Xi\uppi$ decay channel. The extracted scattering parameters for the \LKPlusInt and the \LKMinInt (non-resonant) interaction are in agreement with the femtoscopic measurements of the same pairs performed by the ALICE collaboration in \PbPb collisions.

The presented data provide important experimental constraints for low-energy effective theories, aiming at describing the strangeness $S=0$ and $S=-2$ sectors of the meson--baryon interaction, and show the possibility to investigate the role of still not established resonances, such as the \XRes, in hadron--hadron interactions.

\newenvironment{acknowledgement}{\relax}{\relax}
\begin{acknowledgement}
\section*{Acknowledgements}
The ALICE Collaboration is grateful to Prof.~Francesco Giacosa for the extremely valuable guidance on the theoretical aspects and to Dr.~Albert Feijoo for the fruitful discussions.

The ALICE Collaboration would like to thank all its engineers and technicians for their invaluable contributions to the construction of the experiment and the CERN accelerator teams for the outstanding performance of the LHC complex.
The ALICE Collaboration gratefully acknowledges the resources and support provided by all Grid centres and the Worldwide LHC Computing Grid (WLCG) collaboration.
The ALICE Collaboration acknowledges the following funding agencies for their support in building and running the ALICE detector:
A. I. Alikhanyan National Science Laboratory (Yerevan Physics Institute) Foundation (ANSL), State Committee of Science and World Federation of Scientists (WFS), Armenia;
Austrian Academy of Sciences, Austrian Science Fund (FWF): [M 2467-N36] and Nationalstiftung f\"{u}r Forschung, Technologie und Entwicklung, Austria;
Ministry of Communications and High Technologies, National Nuclear Research Center, Azerbaijan;
Conselho Nacional de Desenvolvimento Cient\'{\i}fico e Tecnol\'{o}gico (CNPq), Financiadora de Estudos e Projetos (Finep), Funda\c{c}\~{a}o de Amparo \`{a} Pesquisa do Estado de S\~{a}o Paulo (FAPESP) and Universidade Federal do Rio Grande do Sul (UFRGS), Brazil;
Bulgarian Ministry of Education and Science, within the National Roadmap for Research Infrastructures 2020-2027 (object CERN), Bulgaria;
Ministry of Education of China (MOEC) , Ministry of Science \& Technology of China (MSTC) and National Natural Science Foundation of China (NSFC), China;
Ministry of Science and Education and Croatian Science Foundation, Croatia;
Centro de Aplicaciones Tecnol\'{o}gicas y Desarrollo Nuclear (CEADEN), Cubaenerg\'{\i}a, Cuba;
Ministry of Education, Youth and Sports of the Czech Republic, Czech Republic;
The Danish Council for Independent Research | Natural Sciences, the VILLUM FONDEN and Danish National Research Foundation (DNRF), Denmark;
Helsinki Institute of Physics (HIP), Finland;
Commissariat \`{a} l'Energie Atomique (CEA) and Institut National de Physique Nucl\'{e}aire et de Physique des Particules (IN2P3) and Centre National de la Recherche Scientifique (CNRS), France;
Bundesministerium f\"{u}r Bildung und Forschung (BMBF) and GSI Helmholtzzentrum f\"{u}r Schwerionenforschung GmbH, Germany;
General Secretariat for Research and Technology, Ministry of Education, Research and Religions, Greece;
National Research, Development and Innovation Office, Hungary;
Department of Atomic Energy Government of India (DAE), Department of Science and Technology, Government of India (DST), University Grants Commission, Government of India (UGC) and Council of Scientific and Industrial Research (CSIR), India;
National Research and Innovation Agency - BRIN, Indonesia;
Istituto Nazionale di Fisica Nucleare (INFN), Italy;
Japanese Ministry of Education, Culture, Sports, Science and Technology (MEXT) and Japan Society for the Promotion of Science (JSPS) KAKENHI, Japan;
Consejo Nacional de Ciencia (CONACYT) y Tecnolog\'{i}a, through Fondo de Cooperaci\'{o}n Internacional en Ciencia y Tecnolog\'{i}a (FONCICYT) and Direcci\'{o}n General de Asuntos del Personal Academico (DGAPA), Mexico;
Nederlandse Organisatie voor Wetenschappelijk Onderzoek (NWO), Netherlands;
The Research Council of Norway, Norway;
Commission on Science and Technology for Sustainable Development in the South (COMSATS), Pakistan;
Pontificia Universidad Cat\'{o}lica del Per\'{u}, Peru;
Ministry of Education and Science, National Science Centre and WUT ID-UB, Poland;
Korea Institute of Science and Technology Information and National Research Foundation of Korea (NRF), Republic of Korea;
Ministry of Education and Scientific Research, Institute of Atomic Physics, Ministry of Research and Innovation and Institute of Atomic Physics and University Politehnica of Bucharest, Romania;
Ministry of Education, Science, Research and Sport of the Slovak Republic, Slovakia;
National Research Foundation of South Africa, South Africa;
Swedish Research Council (VR) and Knut \& Alice Wallenberg Foundation (KAW), Sweden;
European Organization for Nuclear Research, Switzerland;
Suranaree University of Technology (SUT), National Science and Technology Development Agency (NSTDA), Thailand Science Research and Innovation (TSRI) and National Science, Research and Innovation Fund (NSRF), Thailand;
Turkish Energy, Nuclear and Mineral Research Agency (TENMAK), Turkey;
National Academy of  Sciences of Ukraine, Ukraine;
Science and Technology Facilities Council (STFC), United Kingdom;
National Science Foundation of the United States of America (NSF) and United States Department of Energy, Office of Nuclear Physics (DOE NP), United States of America.
In addition, individual groups or members have received support from:
European Research Council, Strong 2020 - Horizon 2020 (grant nos. 950692, 824093), European Union;
Academy of Finland (Center of Excellence in Quark Matter) (grant nos. 346327, 346328), Finland.

\end{acknowledgement}

\bibliographystyle{utphys}   
\bibliography{bibliography}
\newpage
\appendix

%
%

\section{The ALICE Collaboration}
\label{app:collab}
\begin{flushleft} 
\small

S.~Acharya\,\orcidlink{0000-0002-9213-5329}\,$^{\rm 127}$, 
D.~Adamov\'{a}\,\orcidlink{0000-0002-0504-7428}\,$^{\rm 87}$, 
A.~Adler$^{\rm 70}$, 
G.~Aglieri Rinella\,\orcidlink{0000-0002-9611-3696}\,$^{\rm 33}$, 
M.~Agnello\,\orcidlink{0000-0002-0760-5075}\,$^{\rm 30}$, 
N.~Agrawal\,\orcidlink{0000-0003-0348-9836}\,$^{\rm 51}$, 
Z.~Ahammed\,\orcidlink{0000-0001-5241-7412}\,$^{\rm 134}$, 
S.~Ahmad\,\orcidlink{0000-0003-0497-5705}\,$^{\rm 16}$, 
S.U.~Ahn\,\orcidlink{0000-0001-8847-489X}\,$^{\rm 71}$, 
I.~Ahuja\,\orcidlink{0000-0002-4417-1392}\,$^{\rm 38}$
% ...

\section*{Affiliation Notes}

$^{\rm I}$ Also at: Max-Planck-Institut f\"{u}r Physik, Munich, Germany\\
$^{\rm II}$ Also at: Italian National Agency for New Technologies, Energy and Sustainable Economic Development (ENEA), Bologna, Italy\\
$^{\rm III}$ Also at: Department of Physics, Tokyo Metropolitan University, Hachioji, Japan\\
$^{\rm IV}$ Also at: Helmholtz Institut f\"ur Strahlen- und Kernphysik and Bethe Center for Theoretical Physics, Universit\"at Bonn, Bonn, Germany\\
$^{\rm V}$ Also at: Department of Applied Physics, Aligarh Muslim University, Aligarh, India\\
$^{\rm VI}$ Also at: Institute of Theoretical Physics, University of Wroclaw, Poland\\
$^{\rm VII}$ Also at: An institution covered by a cooperation agreement with CERN\\

\section*{Collaboration Institutes}

$^{1}$ A.I. Alikhanyan National Science Laboratory (Yerevan Physics Institute) Foundation, Yerevan, Armenia\\
$^{2}$ AGH University of Science and Technology, Cracow, Poland\\
$^{3}$ Bogolyubov Institute for Theoretical Physics, National Academy of Sciences of Ukraine, Kiev, Ukraine\\
$^{4}$ Bose Institute, Department of Physics  and Centre for Astroparticle Physics and Space Science (CAPSS), Kolkata, India\\
$^{5}$ California Polytechnic State University, San Luis Obispo, California, United States\\
$^{6}$ Central China Normal University, Wuhan, China\\
$^{7}$ Centro de Aplicaciones Tecnol\'{o}gicas y Desarrollo Nuclear (CEADEN), Havana, Cuba\\
$^{8}$ Centro de Investigaci\'{o}n y de Estudios Avanzados (CINVESTAV), Mexico City and M\'{e}rida, Mexico\\
$^{9}$ Chicago State University, Chicago, Illinois, United States\\
$^{10}$ China Institute of Atomic Energy, Beijing, China\\
$^{11}$ China University of Geosciences, Wuhan, China\\
$^{12}$ Chungbuk National University, Cheongju, Republic of Korea\\
$^{13}$ Comenius University Bratislava, Faculty of Mathematics, Physics and Informatics, Bratislava, Slovak Republic\\
$^{14}$ COMSATS University Islamabad, Islamabad, Pakistan\\
$^{15}$ Creighton University, Omaha, Nebraska, United States\\
$^{16}$ Department of Physics, Aligarh Muslim University, Aligarh, India\\
$^{17}$ Department of Physics, Pusan National University, Pusan, Republic of Korea\\
$^{18}$ Department of Physics, Sejong University, Seoul, Republic of Korea\\
$^{19}$ Department of Physics, University of California, Berkeley, California, United States\\
$^{20}$ Department of Physics, University of Oslo, Oslo, Norway\\
$^{21}$ Department of Physics and Technology, University of Bergen, Bergen, Norway\\
$^{22}$ Dipartimento di Fisica, Universit\`{a} di Pavia, Pavia, Italy\\
$^{23}$ Dipartimento di Fisica dell'Universit\`{a} and Sezione INFN, Cagliari, Italy\\
$^{24}$ Dipartimento di Fisica dell'Universit\`{a} and Sezione INFN, Trieste, Italy\\
$^{25}$ Dipartimento di Fisica dell'Universit\`{a} and Sezione INFN, Turin, Italy\\
$^{26}$ Dipartimento di Fisica e Astronomia dell'Universit\`{a} and Sezione INFN, Bologna, Italy\\
$^{27}$ Dipartimento di Fisica e Astronomia dell'Universit\`{a} and Sezione INFN, Catania, Italy\\
$^{28}$ Dipartimento di Fisica e Astronomia dell'Universit\`{a} and Sezione INFN, Padova, Italy\\
$^{29}$ Dipartimento di Fisica `E.R.~Caianiello' dell'Universit\`{a} and Gruppo Collegato INFN, Salerno, Italy\\
$^{30}$ Dipartimento DISAT del Politecnico and Sezione INFN, Turin, Italy\\
$^{31}$ Dipartimento di Scienze MIFT, Universit\`{a} di Messina, Messina, Italy\\
$^{32}$ Dipartimento Interateneo di Fisica `M.~Merlin' and Sezione INFN, Bari, Italy\\
$^{33}$ European Organization for Nuclear Research (CERN), Geneva, Switzerland\\
$^{34}$ Faculty of Electrical Engineering, Mechanical Engineering and Naval Architecture, University of Split, Split, Croatia\\
$^{35}$ Faculty of Engineering and Science, Western Norway University of Applied Sciences, Bergen, Norway\\
$^{36}$ Faculty of Nuclear Sciences and Physical Engineering, Czech Technical University in Prague, Prague, Czech Republic\\
$^{37}$ Faculty of Physics, Sofia University, Sofia, Bulgaria\\
$^{38}$ Faculty of Science, P.J.~\v{S}af\'{a}rik University, Ko\v{s}ice, Slovak Republic\\
$^{39}$ Frankfurt Institute for Advanced Studies, Johann Wolfgang Goethe-Universit\"{a}t Frankfurt, Frankfurt, Germany\\
$^{40}$ Fudan University, Shanghai, China\\
$^{41}$ Gangneung-Wonju National University, Gangneung, Republic of Korea\\
$^{42}$ Gauhati University, Department of Physics, Guwahati, India\\
$^{43}$ Helmholtz-Institut f\"{u}r Strahlen- und Kernphysik, Rheinische Friedrich-Wilhelms-Universit\"{a}t Bonn, Bonn, Germany\\
$^{44}$ Helsinki Institute of Physics (HIP), Helsinki, Finland\\
$^{45}$ High Energy Physics Group,  Universidad Aut\'{o}noma de Puebla, Puebla, Mexico\\
$^{46}$ Horia Hulubei National Institute of Physics and Nuclear Engineering, Bucharest, Romania\\
$^{47}$ Indian Institute of Technology Bombay (IIT), Mumbai, India\\
$^{48}$ Indian Institute of Technology Indore, Indore, India\\
$^{49}$ INFN, Laboratori Nazionali di Frascati, Frascati, Italy\\
$^{50}$ INFN, Sezione di Bari, Bari, Italy\\
$^{51}$ INFN, Sezione di Bologna, Bologna, Italy\\
$^{52}$ INFN, Sezione di Cagliari, Cagliari, Italy\\
$^{53}$ INFN, Sezione di Catania, Catania, Italy\\
$^{54}$ INFN, Sezione di Padova, Padova, Italy\\
$^{55}$ INFN, Sezione di Pavia, Pavia, Italy\\
$^{56}$ INFN, Sezione di Torino, Turin, Italy\\
$^{57}$ INFN, Sezione di Trieste, Trieste, Italy\\
$^{58}$ Inha University, Incheon, Republic of Korea\\
$^{59}$ Institute for Gravitational and Subatomic Physics (GRASP), Utrecht University/Nikhef, Utrecht, Netherlands\\
$^{60}$ Institute of Experimental Physics, Slovak Academy of Sciences, Ko\v{s}ice, Slovak Republic\\
$^{61}$ Institute of Physics, Homi Bhabha National Institute, Bhubaneswar, India\\
$^{62}$ Institute of Physics of the Czech Academy of Sciences, Prague, Czech Republic\\
$^{63}$ Institute of Space Science (ISS), Bucharest, Romania\\
$^{64}$ Institut f\"{u}r Kernphysik, Johann Wolfgang Goethe-Universit\"{a}t Frankfurt, Frankfurt, Germany\\
$^{65}$ Instituto de Ciencias Nucleares, Universidad Nacional Aut\'{o}noma de M\'{e}xico, Mexico City, Mexico\\
$^{66}$ Instituto de F\'{i}sica, Universidade Federal do Rio Grande do Sul (UFRGS), Porto Alegre, Brazil\\
$^{67}$ Instituto de F\'{\i}sica, Universidad Nacional Aut\'{o}noma de M\'{e}xico, Mexico City, Mexico\\
$^{68}$ iThemba LABS, National Research Foundation, Somerset West, South Africa\\
$^{69}$ Jeonbuk National University, Jeonju, Republic of Korea\\
$^{70}$ Johann-Wolfgang-Goethe Universit\"{a}t Frankfurt Institut f\"{u}r Informatik, Fachbereich Informatik und Mathematik, Frankfurt, Germany\\
$^{71}$ Korea Institute of Science and Technology Information, Daejeon, Republic of Korea\\
$^{72}$ KTO Karatay University, Konya, Turkey\\
$^{73}$ Laboratoire de Physique des 2 Infinis, Ir\`{e}ne Joliot-Curie, Orsay, France\\
$^{74}$ Laboratoire de Physique Subatomique et de Cosmologie, Universit\'{e} Grenoble-Alpes, CNRS-IN2P3, Grenoble, France\\
$^{75}$ Lawrence Berkeley National Laboratory, Berkeley, California, United States\\
$^{76}$ Lund University Department of Physics, Division of Particle Physics, Lund, Sweden\\
$^{77}$ Nagasaki Institute of Applied Science, Nagasaki, Japan\\
$^{78}$ Nara Women{'}s University (NWU), Nara, Japan\\
$^{79}$ National and Kapodistrian University of Athens, School of Science, Department of Physics , Athens, Greece\\
$^{80}$ National Centre for Nuclear Research, Warsaw, Poland\\
$^{81}$ National Institute of Science Education and Research, Homi Bhabha National Institute, Jatni, India\\
$^{82}$ National Nuclear Research Center, Baku, Azerbaijan\\
$^{83}$ National Research and Innovation Agency - BRIN, Jakarta, Indonesia\\
$^{84}$ Niels Bohr Institute, University of Copenhagen, Copenhagen, Denmark\\
$^{85}$ Nikhef, National institute for subatomic physics, Amsterdam, Netherlands\\
$^{86}$ Nuclear Physics Group, STFC Daresbury Laboratory, Daresbury, United Kingdom\\
$^{87}$ Nuclear Physics Institute of the Czech Academy of Sciences, Husinec-\v{R}e\v{z}, Czech Republic\\
$^{88}$ Oak Ridge National Laboratory, Oak Ridge, Tennessee, United States\\
$^{89}$ Ohio State University, Columbus, Ohio, United States\\
$^{90}$ Physics department, Faculty of science, University of Zagreb, Zagreb, Croatia\\
$^{91}$ Physics Department, Panjab University, Chandigarh, India\\
$^{92}$ Physics Department, University of Jammu, Jammu, India\\
$^{93}$ Physics Program and International Institute for Sustainability with Knotted Chiral Meta Matter (SKCM2), Hiroshima University, Hiroshima, Japan\\
$^{94}$ Physikalisches Institut, Eberhard-Karls-Universit\"{a}t T\"{u}bingen, T\"{u}bingen, Germany\\
$^{95}$ Physikalisches Institut, Ruprecht-Karls-Universit\"{a}t Heidelberg, Heidelberg, Germany\\
$^{96}$ Physik Department, Technische Universit\"{a}t M\"{u}nchen, Munich, Germany\\
$^{97}$ Politecnico di Bari and Sezione INFN, Bari, Italy\\
$^{98}$ Research Division and ExtreMe Matter Institute EMMI, GSI Helmholtzzentrum f\"ur Schwerionenforschung GmbH, Darmstadt, Germany\\
$^{99}$ RIKEN iTHEMS, Wako, Japan\\
$^{100}$ Saga University, Saga, Japan\\
$^{101}$ Saha Institute of Nuclear Physics, Homi Bhabha National Institute, Kolkata, India\\
$^{102}$ School of Physics and Astronomy, University of Birmingham, Birmingham, United Kingdom\\
$^{103}$ Secci\'{o}n F\'{\i}sica, Departamento de Ciencias, Pontificia Universidad Cat\'{o}lica del Per\'{u}, Lima, Peru\\
$^{104}$ Stefan Meyer Institut f\"{u}r Subatomare Physik (SMI), Vienna, Austria\\
$^{105}$ SUBATECH, IMT Atlantique, Nantes Universit\'{e}, CNRS-IN2P3, Nantes, France\\
$^{106}$ Sungkyunkwan University, Suwon City, Republic of Korea\\
$^{107}$ Suranaree University of Technology, Nakhon Ratchasima, Thailand\\
$^{108}$ Technical University of Ko\v{s}ice, Ko\v{s}ice, Slovak Republic\\
$^{109}$ The Henryk Niewodniczanski Institute of Nuclear Physics, Polish Academy of Sciences, Cracow, Poland\\
$^{110}$ The University of Texas at Austin, Austin, Texas, United States\\
$^{111}$ Universidad Aut\'{o}noma de Sinaloa, Culiac\'{a}n, Mexico\\
$^{112}$ Universidade de S\~{a}o Paulo (USP), S\~{a}o Paulo, Brazil\\
$^{113}$ Universidade Estadual de Campinas (UNICAMP), Campinas, Brazil\\
$^{114}$ Universidade Federal do ABC, Santo Andre, Brazil\\
$^{115}$ University of Cape Town, Cape Town, South Africa\\
$^{116}$ University of Houston, Houston, Texas, United States\\
$^{117}$ University of Jyv\"{a}skyl\"{a}, Jyv\"{a}skyl\"{a}, Finland\\
$^{118}$ University of Kansas, Lawrence, Kansas, United States\\
$^{119}$ University of Liverpool, Liverpool, United Kingdom\\
$^{120}$ University of Science and Technology of China, Hefei, China\\
$^{121}$ University of South-Eastern Norway, Kongsberg, Norway\\
$^{122}$ University of Tennessee, Knoxville, Tennessee, United States\\
$^{123}$ University of the Witwatersrand, Johannesburg, South Africa\\
$^{124}$ University of Tokyo, Tokyo, Japan\\
$^{125}$ University of Tsukuba, Tsukuba, Japan\\
$^{126}$ University Politehnica of Bucharest, Bucharest, Romania\\
$^{127}$ Universit\'{e} Clermont Auvergne, CNRS/IN2P3, LPC, Clermont-Ferrand, France\\
$^{128}$ Universit\'{e} de Lyon, CNRS/IN2P3, Institut de Physique des 2 Infinis de Lyon, Lyon, France\\
$^{129}$ Universit\'{e} de Strasbourg, CNRS, IPHC UMR 7178, F-67000 Strasbourg, France, Strasbourg, France\\
$^{130}$ Universit\'{e} Paris-Saclay Centre d'Etudes de Saclay (CEA), IRFU, D\'{e}partment de Physique Nucl\'{e}aire (DPhN), Saclay, France\\
$^{131}$ Universit\`{a} degli Studi di Foggia, Foggia, Italy\\
$^{132}$ Universit\`{a} del Piemonte Orientale, Vercelli, Italy\\
$^{133}$ Universit\`{a} di Brescia, Brescia, Italy\\
$^{134}$ Variable Energy Cyclotron Centre, Homi Bhabha National Institute, Kolkata, India\\
$^{135}$ Warsaw University of Technology, Warsaw, Poland\\
$^{136}$ Wayne State University, Detroit, Michigan, United States\\
$^{137}$ Westf\"{a}lische Wilhelms-Universit\"{a}t M\"{u}nster, Institut f\"{u}r Kernphysik, M\"{u}nster, Germany\\
$^{138}$ Wigner Research Centre for Physics, Budapest, Hungary\\
$^{139}$ Yale University, New Haven, Connecticut, United States\\
$^{140}$ Yonsei University, Seoul, Republic of Korea\\
$^{141}$  Zentrum  f\"{u}r Technologie und Transfer (ZTT), Worms, Germany\\
$^{142}$ Affiliated with an institute covered by a cooperation agreement with CERN\\
$^{143}$ Affiliated with an international laboratory covered by a cooperation agreement with CERN.\\

\end{flushleft} 

\end{document}